\documentclass[12pt, a4paper,openright]{article}
\usepackage[english]{babel}
\usepackage[T1]{fontenc}
\usepackage{latexsym}
\usepackage{indentfirst}
\usepackage{graphicx}
\usepackage{amsmath,amssymb,amsfonts,amsthm,amstext,amscd}
\usepackage{mathrsfs}
\usepackage{times}
\usepackage{enumerate}
\usepackage[utf8]{inputenc}
\usepackage[mathscr]{eucal}
\usepackage{scrextend}
\usepackage{graphics}% serve per il blocco sotto indicato
\usepackage{titling}

\newtheorem{defi}{Definition}[section]

\newtheorem{lem}{Lemma}[section]
\newtheorem{rem}{Remark}[section]
\newtheorem{thm}{Theorem}[section]

\newtheorem{coro}{Corolary}[section]

\begin{document}
% il blocco che segue serve a tenere il titolo in una sola riga; usa il package graphics
\makeatletter
  \def\title@font{\Large\bfseries}
  \let\ltx@maketitle\@maketitle
  \def\@maketitle{\bgroup%
    \let\ltx@title\@title%
    \def\@title{\resizebox{\textwidth}{!}{%
      \mbox{\title@font\ltx@title}%
    }}%
    \ltx@maketitle%
  \egroup}
\makeatother
% fine blocco
\setlength{\droptitle}{-10em}   % This is my set screw: shifta il titolo
\title{Heisenberg Uncertainty Relations as Statistical Invariants}
\author{Aniello Fedullo \\
\small Department of Physics "E. R. Caianiello", University of Salerno, Italy  \\
\small afedullo@unisa.it}
\date{\vspace{-5ex}}
%\date{}
\maketitle

\begin{addmargin}[16em]{0em}  % 1em left, 2em right
	\small{«Surely, one would like to be able to deduce the quantitative laws of quantum mechanics directly from their
	\emph{anschaulich} foundations, that is, essentially, relation $\delta $p $\delta $q $\sim$ h».}
\end{addmargin}
\begin{addmargin}[21.4em]{0em}
	\small{Werner Heisenberg \cite[p. 196]{heis:grund}}
\end{addmargin}

%\vskip -20pt
\bigskip
\bigskip

\textbf{Abstract.} For a simple set of observables we can express, in terms of transition probabilities alone, the Heisenberg
Uncertainty Relations, so that they are proven to be not only necessary, but sufficient too, in order for the given
observables to admit a quantum model. Furthermore distinguished characterizations of strictly complex and real quantum
models, with some ancillary results, are presented and discussed.

\section{Foreword}
The problem of understanding the empirical basis of the quantum mechanical formalism has been approached, starting from
eighties, by means of the method of statistical invariants \cite{acc:fed} \cite{acc:trends}. The main idea of this approach, borrowed from the
Klein's program of Erlangen \cite{klein:geom}, is to classify the probabilistic models according to statistical invariants, expressed
in terms of the transition probabilities of the physical observables. That is to say, one considers the transition
probabilities as the basic empirical data from which the mathematical model should be deduced. The statistical
invariants for some simple systems were explicitly computed and it was shown that they allowed to distinguish among
Kolmogorovian, real Hilbert space and complex Hilbert space models. Actually necessary and sufficient conditions for
the existence of each model were found \cite{acc:fed} \cite{gudder:models} \cite{fed:exist}. When applied to the quantum-mechanical transition probabilities, they
proved not only the necessity of a non classical probabilistic model, but also the necessity of using complex rather
than real Hilbert spaces \cite{acc:fed} \cite{fed:exist}, so offering the solution to the open problem of ``...singling out in full generality
the empirical basis for the choice of complex numbers in quantum theory....''\cite{jauch:found}. Furthermore the Kolmogorovian
statistical invariant was recognized as a form of the celebrated Bell inequality \cite{acc:fed}, expressed by the transition
probabilities instead of the correlation functions. In the present paper we pursue the study of a triple of
two-dimensional observables undertaken in \cite{acc:fed} and, by means of the notion of quantum models, we show as their
statistical invariants represent the Heisenberg uncertainty relations expressed in terms of transition probabilities
alone. This allow us to affirm that uncertainty relations are not only necessary, but sufficient too, in order for the
given observables to admit a quantum model.

\section{Preliminary definitions and results}

In the oft quoted paper \cite{acc:fed} a triple $A,B,C$ of two-valued observables subject to take values $(a_{\alpha }),(b_{\beta
}),(c_{\gamma })$ were studied, given their transition probabilities
\begin{equation}
	P(A=a_{\alpha }|B=b_{\beta }),P(B=b_{\beta }|C=c_{\gamma }),P(C=c_{\gamma }|A=a_{\alpha })
\end{equation}
under the symmetry assumptions
\begin{equation}
	P(A=a_{\alpha }|B=b_{\beta })=P(B=b_{\beta }|A=a_{\alpha }),...
	\footnote{Here and in the following ellipsis stands for all similar relations involving the observables left over.}
\end{equation}
For observable is meant any quantity arising from experiments of whatever nature. The transition probability
$P(A=a_{\alpha }|B=b_{\beta })$ is the conditional probability that $A$ takes the value $a_{\alpha }$ conditioned by
the fact that $B$ is known to assume the value $b_{\beta }.$
We will denote the transition probabilities (1) as
\begin{equation}
	P(A|B)=\left[\begin{matrix}p&1-p\\1-p&p\end{matrix}\right]=\left[\begin{matrix}\cos ^2\frac{\alpha } 2&\sin
	^2\frac{\alpha } 2\\\sin ^2\frac{\alpha } 2&\cos ^2\frac{\alpha } 2\end{matrix}\right],
\end{equation}
\begin{equation}	
	P(B|C)=\left[\begin{matrix}q&1-q\\1-q&q\end{matrix}\right]=\left[\begin{matrix}\cos
	^2\frac{\beta } 2&\sin ^2\frac{\beta } 2\\\sin ^2\frac{\beta } 2&\cos ^2\frac{\beta } 2\end{matrix}\right],
\end{equation}
\begin{equation}	
	P(C|A)=\left[\begin{matrix}r&1-r\\1-r&r\end{matrix}\right]=\left[\begin{matrix}\cos ^2\frac{\gamma } 2&\sin
	^2\frac{\gamma } 2\\\sin ^2\frac{\gamma } 2&\cos ^2\frac{\gamma } 2\end{matrix}\right],
\end{equation}
assuming, unless otherwise specified, that
\begin{equation}
	0 < p,q,r < 1 ,
\end{equation}
for which the angles can be chosen such that
\begin{equation}
	0 < \alpha ,\beta ,\gamma < \pi.
\end{equation}
The transition probabilities (1) were said to admit a complex (resp. a real) Hilbert space model if there exist three 
orthonormal bases $\{\phi _{\alpha }\},\{\psi _{\beta }\},\{\chi _{\gamma }\}$ of a two-dimensional complex (resp.
real) Hilbert space $\mathcal{H}$ such that 
\begin{equation}
	P(A=a_{\alpha }|B=b_{\beta })=|\langle \phi _{\alpha }|\psi _{\beta }\rangle |^2,...
\end{equation}
In particular a complex Hilbert space model was said a spin model if the three o. n. bases can be taken as the
normalized eigenvectors $\psi _{\alpha }(u_A),\psi _{\beta }(u_B),\psi _{\gamma }(u_C)$  $(\alpha ,\beta ,\gamma =1,2)$
of the spin operators $u_A\cdot \sigma ,u_B\cdot \sigma ,u_C\cdot \sigma $ for some $u_A,u_B,u_C\in S^{(2)},$ where
$S^{(2)}$ denotes the real unit sphere in $\mathbb{R}^3$ and $u\cdot \sigma $ is defined in terms of the Pauli matrices
\begin{equation}
	\sigma _1:=\left[\begin{matrix}0&1\\1&0\end{matrix}\right],\;\;\;\sigma
	_2:=\left[\begin{matrix}0&-i\\i&0\end{matrix}\right],\;\;\;\sigma _3:=\left[\begin{matrix}1&0\\0&-1\end{matrix}\right]
\end{equation}
as
\begin{equation}
	u\cdot \sigma :=u_1\sigma _1+u_2\sigma _2+u_3\sigma _3
\end{equation}
In the latter case we can write (8), in terms of the angles $\widehat {u_Au_B},\widehat {u_Bu_C},\widehat {u_Cu_A},$ as
\begin{equation}
	|\langle \psi _1(u_A)|\psi _1(u_B)\rangle |^2=\cos ^2\frac{\widehat {u_Au_B}} 2,...
\end{equation}
\begin{equation}
	|\langle \psi _1(u_A)|\psi _2(u_B)\rangle |^2=\sin ^2\frac{\widehat {u_Au_B}} 2,...
\end{equation}
In the present paper we prefer focus our attention on observables rather than transition probabilities alone. So we can
recover the usual formalism of quantum mechanics \cite{neum:grund} according to observables are postulated in correspondence with
self-adjoint operators on a suitable complex Hilbert space. In this frame uncertainty relations we are concerned found
their most general formulation, so that we are led to give the following
\begin{defi}
The observables $A,B,C$ are said to admit a quantum model if and only if there exists a complex Hilbert
space $\mathcal{H}$ and self-adjoint operators $\hat A,\hat B,\hat C$ acting\footnote{$\hat A,\hat B,\hat C$ are defined up to
a common unitary transformation, cfr. \cite{acc:fed} corollary 8. A self-adjoint operator having eigenvalues $\mp 1$ is just a
spin operator.} on it such that the values of each observable coincide with the eigenvalues of the corresponding
operator and the transition probabilities (1) admit the complex Hilbert space model defined by the o. n. bases of $\mathcal{H}$
made up of the normalized eigenvectors\footnote{Said eigenstates as well.} of $\hat A,\hat B,\hat C.$ The model will
be called a real quantum model if $\mathcal{H}$ can be taken real, or a strictly complex quantum model otherwise.
\end{defi}
Next theorem shows a first expected link between quantum models of observables and Hilbert space models of the relative
transition probabilities.
\begin{thm}
The following assertions are equivalent:
\ i) the observables $A,B,C$ admit a quantum model;
\ ii) their transition probabilities admit a spin model;
\ iii) their transition probabilities admit a complex Hilbert space model.
\end{thm}
\begin{proof} To prove that i) is equivalent to ii) it is suffices to observe that, for every observable $X$(with values
$x_{1,}x_2\text )$ of the triple $A,B,C,$ the operator
\begin{equation}
	\hat S_X:=\frac 2{x_1-x_2}\hat X-\frac{x_1+x_2}{x_1-x_2}\hat 1, 
\end{equation}
where $\hat 1$ is the identity operator, has the eigenvalues $\mp 1,$ so it is a spin operator and, due to $[\hat
S_X,\hat X]=0,$ it has the same eigenvectors of $\hat X.$ Further, the equivalence between ii) and iii) was proven in
\cite{acc:fed}, theorem 7\footnote{\label{fn:nota}Cfr. Appendix, 1.}. 
\end{proof}
\begin{rem} The particular case of real quantum models will be discussed later. Notice that the precedent result deals
with a (linear) rescaling of the observables preserving probabilities. Further observe that the spin operators $\hat
S_A,\;\hat S_B,\;\hat S_C$ can be written as $u_A\cdot \sigma ,\;u_B\cdot \sigma ,\;u_C\cdot \sigma $ respectively, for
suitable unit vectors\footnote{Any common unitary transformation of $\hat A,\hat B,\hat C,$ referred to in note 2,
induces a common rotation of $u_A,u_B,u_C,$ cfr. \cite{acc:fed}, corollary 8 proof.} $u_A,\;u_B,\;u_C\in S^{(2)},$ which so remain
associated to the self-adjoint operators $\hat A,\;\hat B,\;\hat C$ respectively.
\end{rem}
In this framework we are able to reformulate some results of \cite{acc:fed} as follows:
\begin{thm} The following assertions are equivalent:
i)  the observables $A,B,C$ admit a quantum model;
ii) there exist\footnote{Observe that $u_A,u_B,u_C$ are necessarily distinct and pairwise non collinear,
		due to the assumption (5).} $u_A,u_B,u_C\in S^{(2)}$ such that $u_A\cdot u_B=\cos \alpha ,\;u_B\cdot u_C=\cos \beta
		,\;u_C\cdot u_A=\cos \gamma ;$
iii) $1-\cos ^2\alpha -\cos ^2\beta -\cos ^2\gamma +2\;\cos \alpha \;\cos \beta \;\cos \gamma \ge 0$.
Moreover inequalities iii) is saturated if and only if $A,B,C$ admit a real quantum model; in such a
case, and only then, the unit vectors in ii) are coplanar.
\end{thm}
\begin{proof} By theorem 2.1, i) is equivalent to the existence of a spin model for the transition probabilities; this in
turn, by eqs. (8), (11) and (12), as well as due to $\cos \widehat {u_Au_B}=u_A\cdot u_B,\cos \widehat {u_Bu_C}=u_B\cdot
u_C,\widehat {u_Au_C}=u_A\cdot u_C,$ is easily recognized equivalent to ii); the equivalence between ii) and iii) was
established in \cite{acc:fed}\textsuperscript{\ref{fn:nota}}, proposition 3, just as, from theorems 9 and 10 therein, it follows straight the last
statement to be proven.
\end{proof}
\begin{rem} The inequality in iii) of the precedent theorem is said a statistical invariant \cite{acc:trends} for a quantum model. It
is said in particular a statistical invariant for a real quantum model if the inequality is saturated, or a statistical
invariant for a strictly complex quantum model, otherwise. Many equivalent forms of these invariants were discovered in
\cite{acc:fed}\textsuperscript{\ref{fn:nota}} and some others, involving uncertainty relations, will appear below.
\end{rem}
\section{Uncertainty Relations}
Uncertainty relations were introduced in quantum mechanics by Heisenberg \cite{heis:grund} and successively extended and strengthen by
many authors \cite{rob:unc}\cite{schr:zum}\footnote{For a recent review cfr. \cite{sen:unc} and the bibliography therein.}. In this paper we will
refer to the following stronger form:
\begin{thm} (Schr\"odinger \cite{schr:zum}) For every couple of self-adjoint operators $\hat X,\hat Y$ acting on a complex 
Hilbert space $\mathcal{H}$  and for every state\footnote{As known a state is defined as a norm 1 element of $\mathcal{H}$ up to a phase
factor.} $\psi $ the following inequality holds\footnote{The first addend in the r. h. s. is said the \emph{covariance} term
and the second the \emph{commutator} term.}:
\begin{equation}
	\;\mathit{Var}(\hat X)\mathit{Var}(\hat Y)\ge (\frac 1 2\langle \{\hat X,\hat Y\}\rangle -\langle \hat X\rangle
	\langle \hat Y\rangle )^2+(\frac 1{2i}\langle [\hat X,\hat Y]\rangle )^2
\end{equation}
where $\langle \hat Z\rangle :=\langle \psi |\hat Z|\psi \rangle $ and $\mathit{Var}(\hat Z):=\langle \widehat 
Z^2\rangle -\langle \widehat  Z\rangle ^2$ are resp. the average of $\hat Z$ and the \ \ \ \ variance of $\hat Z$ in
the state $\psi ,$ with $[\hat X,\hat Y]:=\hat X\hat Y\;-\;\hat Y\hat X$ and $\{\hat X,\hat Y\}:=\hat X\hat Y\;+\;\hat
Y\hat X$ being the commutator, resp. the anticommutator, of $\hat X$ and $\hat Y.$
\end{thm}
\begin{proof} Cfr. for example reference \cite{grif:quan} and, of course, \cite{schr:zum}.
\end{proof}
\begin{rem} In the following lemma 4.4 we show that, for the operators we consider in this paper, the inequality expressing
the Heisenberg-Schr\"odinger uncertainty relation is in fact saturated, so that it assumes the form of an identity.
From now on $\mathcal{H}$ will denote a two-dimensional complex Hilbert space.
\end{rem}
\section{Some useful lemmas}
\begin{lem} Whichever $\hat X,\hat Y,\hat Z$ are taken among the operators $\hat A,\hat B,\hat C$ associated to the 
given observables $A,B,C$ one has, for every state, 
\begin{equation}
	\langle \hat Z\rangle =\frac{z_1+z_2} 2+\frac{z_1-z_2} 2\;\langle u_Z\cdot \sigma \rangle
\end{equation}
and
\begin{equation}
	\mathit{Var}(\hat Z)=(\frac{z_1-z_2} 2)^2\mathit{Var}(u_Z\cdot \sigma ) 
\end{equation}
where $z_{1,}z_2$ are the values of the observable  $Z,$ as well as
\begin{equation}
	\frac 1 2\langle \{\hat X,\hat Y\}\rangle -\langle \hat X\rangle \langle \hat Y\rangle =\frac{x_1-x_2}
	2\frac{y_1-y_2} 2\;\;(\frac 1 2\langle \{u_X\cdot \sigma ,u_Y\cdot \sigma \}\rangle -\langle u_X\cdot \sigma \rangle
	\langle u_Y\cdot \sigma \rangle ) 
\end{equation}
\ and
\begin{equation}
	\frac 1{2i}\langle [\hat X,\hat Y]\rangle =\frac{x_1-x_2} 2\frac{y_1-y_2} 2\;\;\frac
	1{2i}\langle [u_X\cdot \sigma ,u_Y\cdot \sigma ]\rangle.
\end{equation}
Consequently the Heisenberg-Schr\"odinger uncertainty relation (14) holds for a couple of
operators $\hat X,\hat Y$ acting on $\mathcal{H}$ if and only if it holds for the associated spin
operators $u_X\cdot \sigma ,\;\;u_Y\cdot \sigma .$ Furthermore the former relation is saturated if and
only if the latter is.
\end{lem}
\begin{proof} Cfr. Appendix, 2.
\end{proof}
\begin{lem} For every $u,v,w\in S^{(2)}$ the following identities hold in each eigenstate $\psi _k(w)$ of $w\cdot \sigma $  $(k=1,2):$ 
\begin{equation}
	\langle u\cdot \sigma \rangle =(-1)^{k-1}u\cdot w,
\end{equation}
\begin{equation}
	\mathit{Var}(u\cdot \sigma )=1-(u\cdot w)^2,
\end{equation}
\begin{equation}
	\frac 1{2i}\langle [u\cdot \sigma ,v\cdot \sigma ]\rangle =(u\times v)\cdot w
\end{equation}
and just for every state:
\begin{equation}
	\frac 1 2\langle \{u\cdot \sigma ,v\cdot \sigma \}\rangle =u\cdot v.
\end{equation}
\end{lem}
\begin{proof} Cfr. Appendix, 3
\end{proof}
\begin{lem} For every state $\psi $ of the Hilbert space $\mathcal{H}$ there is a $w\in S^{(2)}$ such that
$\psi =\psi _1(w),$ where $\psi _1(w)$ is the eigenstate\footnote{One has also that $\psi =\psi _2(-w)$, but it is
not required here; $w$ is known as a representation of the state $\psi $ on the Bloch's sphere $S^{(2)}$.} of the
spin operator $w\cdot \sigma ,$ corresponding to the eigenvalue $1.$
\end{lem}
\begin{proof} We can put, up to an non influential phase factor, $\psi =\left[\begin{matrix}\left|\psi _1\right|&\Re (\psi
_2)+i\Im (\psi _2)\end{matrix}\right]^T$ while as known\footnote{In a basis of $\mathcal{H}$ in which the Pauli matrices have the usual form (9), cfr. for example \cite{acc:fed}, p. 170.}
$\psi _1(w)=\left[\begin{matrix}\sqrt{\frac{1+w_3} 2}&\frac{w_1+i\;w_2}{\sqrt{2(1+w_3)}}\end{matrix}\right]^T,$ so that
we can solve the vector equation $\psi _1(w)=\psi $ for $w,$ getting $w_1=2\left|\psi _1\right|\;\Re (\psi _2),$ 
$w_2=2\left|\psi _1\right|\;\Im (\psi _2),$  $w_3=2\left|\psi _1\right|^2-1.$ 
\end{proof}
\begin{lem} For every couple of self-adjoint operators $\hat X,\hat Y$ on a two-dimensional complex Hilbert
space $\mathcal{H}$  and for every state $\psi $ the following identity holds:
\begin{equation}
	\;\mathit{Var}(\hat X)\mathit{Var}(\hat Y)=(\frac 1 2\langle \{\hat X,\hat Y\}\rangle -\langle \hat X\rangle
	\langle \hat Y\rangle )^2+(\frac 1{2i}\langle [\hat X,\hat Y]\rangle )^2
\end{equation}
\end{lem}
\begin{proof} Due to lemma 4.1 we can limit ourselves to the case in which $\hat X,\hat Y$ are spin operators, that is $\hat
X=u\cdot \sigma $ and $\hat Y=v\cdot \sigma $ for suitable $u,v\in S^{(2)}.$ However the generic state $\psi ,$ by
lemma 4.3, can be written as $\psi =\psi _1(w),$ for a suitable $w\in S^{(2)}.$ So (23) by means of lemma 4.2 can be
written as
\begin{equation}
	(1-(u\cdot w)^2)(1-(v\cdot w)^2)=(u\cdot v-(u\cdot w)(v\cdot w))^2+((u\times v)\cdot w)^2;
\end{equation}
expanding and simplifying it becomes
\begin{equation}
	1-(u\cdot w)^2-(v\cdot w)^2-(u\cdot v)^2+2(u\cdot v)(u\cdot w)(v\cdot w)=((u\times v)\cdot w)^2 
\end{equation}
which is surely identically satisfied, since each side equals the square of the volume of the parallelepiped of sides
$u,v,w.$  
\end{proof}
\section{The main result}
\begin{thm} Assuming that the two-valued observables $A,B,C$ admit a quantum model, the following
assertions\footnote{Similar assertions hold taking any permutation of the operators $\hat A,\hat B,\hat C.$ } hold
and are equivalent:
	i) $\hat A,\hat B$ satisfy the saturated Heisenberg-Schr\"odinger uncertainty relation for every state:
\begin{equation}
	\;\mathit{Var}(\hat A)\mathit{Var}(\hat B)=(\frac 1 2\langle \{\hat A,\hat B\}\rangle
	-\langle \hat A\rangle \langle \hat B\rangle )^2+(\frac 1{2i}\langle [\hat A,\hat B]\rangle )^2;
\end{equation}
	ii)  the following inequality\footnote{where $\Delta \hat Z:=\sqrt{\mathit{Var}(\hat Z)}$ denotes the
\emph{standard deviation} of $\hat Z.$ The r. h. s. is said the \emph{correlation} term.} holds in each eigenstate $\psi _k(C)$ of
$C$  $(k=1,2):$ 
\begin{equation}
	\Delta A\;\Delta B\;\ge \;\left|\frac 1 2\langle \{\hat A,\hat B\}\rangle -\langle
	\hat A\rangle \langle \hat B\rangle \right|; 
\end{equation}
	iii)  the following inequality holds:
\begin{equation}
	4pqr-(p+q+r-1)^2\ge 0.
\end{equation}
Furthermore iii) implies the hypothesis and the previous inequalities are saturated if and only if $A,B,C$ admit a real quantum model.
\end{thm}
\begin{proof} i) follows from the hypothesis by lemma 4.4, since the operators associated to $A,B,C$ are self-adjoint, acting
on a two-dimensional complex Hilbert space, by definition of quantum model. If i) holds for every state then, in
particular, it shall hold in each of the two eigenstates of $C;$ so, omitting the last addend and taking the square
roots, one gets ii). By lemma 4.1, we can replace the operators $\hat A,\hat B,\hat C$ with the corresponding spin
operators $u_A\cdot \sigma ,\;\;u_B\cdot \sigma ,\;\;u_C\cdot \sigma ,$ achieving in each of the two eigenstates $\psi
_k(u_C)$  
\begin{equation}
	\Delta (u_A\cdot \sigma )\;\Delta (u_B\cdot \sigma )\;\ge \left|\frac 1 2\langle 
	\{u_A\cdot \sigma ,u_B\cdot \sigma \}\rangle -\langle u_A\cdot \sigma \rangle \langle u_B\cdot \sigma \rangle \right|;
\end{equation}
so the difference between the squares of l. h. s.  and the r. h. s. must satisfy
\begin{equation}
	\mathit{Var}(u_A\cdot \sigma )\mathit{Var}(u_B\cdot \sigma )\;-\ (\frac 1 2\langle \{u_A\cdot
	\sigma ,u_B\cdot \sigma \}\rangle -\langle u_A\cdot \sigma \rangle \langle u_B\cdot \sigma \rangle )^2\ge 0
\end{equation}
that, taking account of (24) and (25) in the proof of lemma 4.4, can be written
\begin{equation}
	1-(u_A\cdot u_B)^2-(u_B\cdot u_C)^2-(u_C\cdot u_A)^2+2(u_A\cdot u_B)(u_B\cdot u_C)(u_C\cdot u_A)\ge 0 
\end{equation}
namely, since $\cos \widehat {u_Au_B}=u_A\cdot u_B,\;\cos \widehat {u_Bu_C}=u_B\cdot u_C,\;\widehat {u_Cu_A}=u_C\cdot u_A,$ 
\begin{equation}
	1-\cos ^2\alpha -\cos ^2\beta -\cos ^2\gamma +2\;\cos \alpha \;\cos \beta \;\cos \gamma \ge 0
\end{equation}
and this by iii) of theorem 2.2 implies the hypothesis. Further (32) is also equivalent to iii) because, with the
notations of section 2, its l. h. s. equals\footnote{Cfr. Appendix, 4.} $4(4pqr-(p+q+r-1)^2)$. Finally the saturation
statement follows from the last part of theorem 2.2.
\end{proof}
\begin{rem} 
	It appears quite astonishing that the, seemingly, much more weak relation in ii) above turns out rather to be
	equivalent to the Heisenberg-Schr\"odinger uncertainty relation in full generality.
\end{rem}
\section{Ancillary results}
\begin{coro}  Under the hypothesis of the preceding theorem, the observables $A,B,C$ admit a strictly complex 
quantum model, resp. a real quantum model, if and only if\footnote{Cfr. Note 12. The symbol $\langle \;\;
\rangle _{\psi }$ denotes the average computed in the state $\psi .$ } $\langle [\hat A,\hat B]\rangle
_{\psi _k(C)}\neq 0,$ resp. $\langle [\hat A,\hat B]\rangle _{\psi _k(C)}=0,$ in each eigenstate $\psi _k(C)$ 
of $C$ $(k=1,2)$.
\end{coro}
\begin{proof} It has been shown in the proof of the preceding theorem that the assertion ii) is equivalent to the
inequality (31), whose l. h. s., taking account of (25), equals $((u_A\times u_B)\cdot u_C)^2$ that, in turn, due to
eq. (21) of lemma 4.2, equals $(\frac 1{2i}\langle [u_A\cdot \sigma ,u_B\cdot \sigma ]\rangle _{\psi _1(u_C)})^2.$
Therefore the assertion ii) of the theorem, thanks to lemma 4, turns out to be equivalent to $(\langle [\hat A,\hat
B]\rangle _{\psi _k(C)})^2\ge 0.$ Moreover it will be saturated as soon as (27) will be and this completes the proof.  
\end{proof}
\begin{coro} If the observables $A,B,C$ admit a strictly complex quantum model then every couple of the
associated operators do not commute. Further, for every state, at least two couples have non vanishing
commutators averages.
\end{coro}
\begin{proof} Let $X,Y,Z$ be whichever permutation of $A,B,C.$ The assumption $[\hat X,\hat Y]=0$ would imply $\langle
[\hat X,\hat Y]\rangle _{\psi _1(Z)}=0,$ against what asserted in the preceding corollary 6.1, and this proves the first
statement. Further, by lemma 4.4, for every state $\psi $ there is a unit vector $w\in S^{(2)}$ such that, due to eq.
(21) of lemma 4.2, we can write
\begin{equation}
	\frac 1{2i}\langle [u_X\cdot \sigma ,u_Y\cdot \sigma ]\rangle _{\psi }=(u_X\times u_Y)\cdot w.
\end{equation}
If this is zero then $w$ belongs to the plane spanned by $u_X,u_Y$ and may be neither in the plane $u_Y,u_Z$ nor in the
plane $u_Z,u_X,$ because $u_A,u_B,u_C$ are not coplanar\footnote{Furthermore cfr. note 6. } by theorem 2.2. So we get 
$\frac 1{2i}\langle [u_Y\cdot \sigma ,u_Z\cdot \sigma ]\rangle _{\psi }\neq 0.$ and $\frac 1{2i}\langle [u_Z\cdot
\sigma ,u_X\cdot \sigma ]\rangle _{\psi }\neq 0.$ and, by eq. (18) of lemma 4.1, the last statement is proven as well.
\end{proof}
\section{Conclusions}
We have proven that the statistical invariant $4pqr-(p+q+r-1)^2\ge 0$ is the expression of the Heisenberg-Schr\"odinger
uncertainty relations for every couple of observables of the considered system. It depends neither on the values of the
observables nor on their scales and units of measure but only on the transition probabilities and provide a condition
not only necessary, but sufficient too, in order for a quantum model to exist. In particular the inequality is strict
if and only if there exists a strictly complex quantum model. In this case some ancillary results involving commutators
have been found. Furthermore real quantum models have been characterized by the saturation  of the uncertainty relation
ii) in theorem 5.1 or, alternatively, by the vanishing of the commutator average appearing in corollary 6.1 or,
definitively, in terms of transition probabilities alone, by the equation $4pqr-(p+q+r-1)^2=0.$ The latter confirms
however the exceptional character of real quantum models, requiring an unlikely functional dependence of the given
transition probabilities. In closing, it is to highlight that, as the transition probabilities can be estimated
starting from relative frequencies experimentally observed, we are able in principle, for the considered simple system,
to deduce the mathematical quantum formalism from the Heisenberg uncertainty relations alone.

\section*{Appendix}
1. Results quoted from \cite{acc:fed}.

\ \ \textit{Theorem 7.} The following assertions are equivalent:

\ \ \ \   i)  the transition matrices $P,Q,R$ admit a complex Hilbert space model;

\ \ \ \  ii)  the transition matrices $P,Q,R$ admit a spin model;

\ \ \ \ iii) $\cos ^2\alpha +\cos ^2\beta +\cos ^2\gamma -1\le 2\;\cos \alpha \;\cos \beta \;\cos \gamma ;$ 

\ \ \ \ iv) $-1\le \frac{\cos ^2\frac{\alpha } 2+\cos ^2\frac{\beta } 2+\cos ^2\frac{\gamma } 2\;-\;1}{2\cos
\frac{\alpha } 2\cos \frac{\beta } 2\cos \frac{\gamma } 2}\le 1;$ 

\ \ \ \  v) $-1\;\le \;\frac{p\;+\;q\;+\;r\;-\;1}{2\sqrt{p\;q\;r}}\;\le \;1;$ 

\ \ \ \ vi) $[\sqrt{pq}-\sqrt{(1-p)(1-q)}]^2\;\le \;r\;\le \;[\sqrt{pq}+\sqrt{(1-p)(1-q)}]^2$ \\

\ \ \textit{Proposition 3} Tree vectors $a,b,c\in S^{(2)}$ satisfying

\begin{equation*}
	\cos \alpha =\cos \widehat {ab},\;\cos \beta =\cos \widehat {bc},\;\cos \gamma =\cos \widehat {ca}
\end{equation*}
\ \ \ \ exist if and only if

\begin{equation*}
	\cos ^2\alpha +\cos ^2\beta +\cos ^2\gamma -1\le 2\;\cos \alpha \;\cos \beta \;\cos \gamma .
\end{equation*} \\
 
\textit{Theorem 9.} The transition matrices $P,Q,R$ admit a real Hilbert space model if and only if

 $\frac{p\;+\;q\;+\;r\;-\;1}{2\sqrt{p\;q\;r}}\;=\;+1$ or $\frac{p\;+\;q\;+\;r\;-\;1}{2\sqrt{p\;q\;r}}\;=\;-1$ 

\ or equivalently

 $\sqrt r\;=\;\sqrt{pq}+\sqrt{(1-p)(1-q)}$ or  $\sqrt r\;=\;\left|\sqrt{pq}-\sqrt{(1-p)(1-q)}\right|.$  

\ \ \textit{Theorem 10.} The transition matrices $P,Q,R$ admit a real Hilbert space model if and only if they admit a
spin model defined by a coplanar triple of vectors in $S^{(2)}.$  \\

2. Proof of lemma 4.1.

Put for simplicity $z_{+\ }:=\frac{z_1+z_2} 2$ and $z_{-\ }:=\frac{z_1-z_2} 2,$ by equation (10) we get

\begin{equation*}
	\langle \hat Z\rangle =z_{+\ }+z_{-\ }\langle u_Z\cdot \sigma \rangle 
\end{equation*}
and, since $\langle (u_Z\cdot \sigma )^2\rangle =\langle \hat 1\rangle =1,$ 

\begin{equation*}
	\langle \hat Z^2\rangle =(z^2)_{+\ }+2z_{+\ }z_{-\ }\langle u_Z\cdot \sigma \rangle ,
\end{equation*}
from which we obtain 

\begin{equation*}
	\mathit{Var}(\hat Z)=z^2_{-{\ }}\mathit{Var}(u_Z\cdot \sigma ).
\end{equation*}
Further, with easily understood notations, we have

\begin{equation*}
	\frac 1 2\langle \{\hat X,\hat Y\}\rangle \;=\;x_{+\ }y_{+\ }+\;x_{-\ }y_{+\ }\langle u_X\cdot \sigma \rangle
	\;+\;x_{+\ }y_{-\ }\langle u_Y\cdot \sigma \rangle \;+\;\frac 1 2\langle \{u_X\cdot \sigma ,u_Y\cdot \sigma \}\rangle ,
\end{equation*}
so that

\begin{equation*}
	\frac 1 2\langle \{\hat X,\hat Y\}\rangle -\langle \hat X\rangle \langle \hat Y\rangle \;=\;x_{-\ }y_{-\ }(\frac 1
	2\langle \{u_X\cdot \sigma ,u_Y\cdot \sigma \}\rangle -\langle u_X\cdot \sigma \rangle \langle u_Y\cdot \sigma \rangle);
\end{equation*}
last, quite directly, we get

\begin{equation*}
	\frac 1{2i}\langle [\hat X,\hat Y]\rangle \;=\;x_{-\ }y_{-\ }\frac 1{2i}\langle [u_X\cdot \sigma ,u_Y\cdot \sigma]\rangle .
\end{equation*}
\begin{flushright}
	$\Box$
\end{flushright}

3 Proof of lemma 4.2. By definition $u\cdot \sigma =\left[\begin{matrix}u_3&u_1-iu_2\\u_1+iu_2&-u_3\end{matrix}\right],$
so that for every state $\psi $ one has

\begin{equation*}
	\langle u\cdot \sigma \rangle =\langle (u\cdot \sigma )\psi |\psi \rangle =[\overline{\psi _1}\;\;\;\overline{\psi
	_2}][u_3\psi _1+(u_1-iu_2)\psi _2\;\;\;\;\;(u_1+iu_2)\psi _1-u_3\psi _2]^T
\end{equation*}
\begin{equation*}
	\text =2u_1\Re (\overline{\psi _1}\psi _2)+2u_2\Im (\overline{\psi _1}\psi _2)+u_3(\left|\psi _1\right|^2-\left|\psi
	_2\right|^{2.}).
\end{equation*}
Since\footnote{In a basis of $\mathcal{H}$ in which the Pauli matrices have the usual form (9), cfr. for example \cite{acc:fed}, p. 170.}
as known $\psi _1(w)=\left[\begin{matrix}\sqrt{\frac{1+w_3} 2}&\frac{w_1+i\;w_2}{\sqrt{2(1+w_3)}}\end{matrix}\right]^T$
and $\psi _2(w)=\left[\begin{matrix}\sqrt{\frac{1-w_3} 2}&-\frac{w_1+i\;w_2}{\sqrt{2(1-w_3)}}\end{matrix}\right]^T,$ 
putting them in the former formula, with easy calculations we get $\langle u\cdot \sigma \rangle _{\psi _1(w)}=u\cdot w$
and $\langle u\cdot \sigma \rangle _{\psi _2(w)}=-u\cdot w$ as asserted in eq. (19). Further it is soon seen that
$(u\cdot \sigma )^2=\hat 1,$ so that $\langle (u\cdot \sigma )^2\rangle =1,$ for which, in the said states,
$\mathit{Var}(u\cdot \sigma )=\langle u\cdot \sigma \rangle ^2-(\langle u\cdot \sigma \rangle )^2=1-(u\cdot w)^2$ as
stated in eq. (20), that therefore is proven. Further, due to\footnote{$\varepsilon _{\mathit{jkl}}$ denotes the
Levi-Civita symbol; on repeated indices summation is understood.} $[\sigma _j,\sigma _k]=2i\varepsilon
_{\mathit{jkl}}\sigma _l$ for every $j,k,$ one has $\frac 1{2i}[u\cdot \sigma ,v\cdot \sigma
]=\mathit{Det}\left[\begin{matrix}\sigma _1&\sigma _2&\sigma _3\\u_1&u_2&u_3\\v_1&v_2&v_3\end{matrix}\right]=(u\times
v)\cdot \sigma ,$ so that, taking the averages in the states $\psi _k(w)$ and considered that $\langle \sigma _k\rangle
=w_k$ for $k=1,2,3$, (21) is proven. Lastly, due to $\{\sigma _h,\sigma _k\}=2\delta _{\mathit{hk}}\hat 1$ for every
$h,k,$ we have $\{\sigma \cdot u,\sigma \cdot v\}=2u\cdot v\hat 1$ so that, taking the averages in whichever state, we
get eq. (22) and the proof of the lemma is complete.

\begin{flushright}
	$\Box$
\end{flushright}

4. With the notations of section 2, thanks to the trigonometric identity $\cos \theta =2\cos ^2\frac{\theta } 2-1,$ we
can write $\cos \alpha =2p-1,\;\cos \beta =2q-1,\;\cos \gamma =2r-1,$ so that

\begin{equation*}
  \begin{gathered}
	1-\cos ^2\alpha -\cos ^2\beta -\cos ^2\gamma +2\;\cos \alpha \;\cos \beta \;\cos \gamma \text
	=\\1-(2p-1)^2-(2q-1)^2-(2r-1)^2+2(2p-1)(2q-1)(2r-1)
  \end{gathered}	
\end{equation*}
that suitably simplified becomes $4(4pqr-(p+q+r-1)^2)$ as asserted.
\begin{flushright}
	$\Box$
\end{flushright}


\begin{thebibliography}{99}
	\providecommand{\natexlab}[1]{#1}
	\providecommand{\url}[1]{\texttt{#1}}
	\expandafter\ifx\csname urlstyle\endcsname\relax
	\providecommand{\doi}[1]{doi: #1}\else
	\providecommand{\doi}{doi: \begingroup \urlstyle{rm}\Url}\fi
	
	\bibitem{acc:fed}
	Accardi L. and A. Fedullo (1982),
	On the statistical Meaning of Complex Numbers in Quantum Mechanics,
	\textit{Lettere al Nuovo Cimento} \textbf{34} 7 161-172.
	
	\bibitem{acc:trends}
	Accardi L. (1984),
	Some trends and problems in quantum probability, in \textit{Quantum Probability and applications to the quantum
	theory of irreversible processes}, L. Accardi, A. Frigerio and V. Gorini (eds.), Springer LNM 1055 1-19.
	
	\bibitem{klein:geom}
	Klein F. (1893),
	Vergleichende Betrachtungen \"uber neuere geometrische Forschungen,
	\textit{Mathematische Annalen} \textbf{43} 63-100. Gesammelte Abh., Springer, 1 (1921) 460-497).
	English translation: A comparative review of recent researches in geometry, by Mellen Haskell, \textit{Bull. N. Y. Math. Soc}
	\textbf{2} (1892-1893) 215-249.
	
	\bibitem{gudder:models}
	Gudder S. and Zangh\`i N. (1982),
	Probability Models,
	\textit{Il Nuovo Cimento} \textbf{79} B 2 291-301.
	
	\bibitem{fed:exist}
	Fedullo A. (1992),
	On the Existence of a Hilbert Space Model for Finite Valued Observables,
	\textit{Il Nuovo Cimento} \textbf{107} B 12 1413-1426. 
	
	\bibitem{jauch:found}
	Jauch J. M. (1968),
	\textit{Foundations of Quantum Mechanics},
	Addison-Wesley Publishing Company, 1968.
	
	\bibitem{neum:grund}
	von Neumann J. (1932), 
	\textit{Mathematische Grundlagen der Quantenmechanik}, Die Grundlehren der Mathematischen Wissenschaften,
	Band 38, Berlin, Springer.
	English translation: \textit{Mathematical Foundations of Quantum Mechanics},
	Princeton University Press, 1971.
	
	\bibitem{heis:grund}
	Heisenberg, W. (1927),
	Uber den anschaulichen Inhalt der quantentheoretischen Kinematik und Mechanik,
	\textit{Zeitschrift f\"ur Physik} \textbf{43} (3-4) 172-198.
	English translation in \cite[62–84]{whi:zur}
	
	\bibitem{sen:unc}
	Sen D. (2014),
	The uncertainty relations in quantum mechanics,
	\textit{Current Science} \textbf{107} (2) 203-218.

	\bibitem{rob:unc}
	Robertson, H.P. (1929),
	The uncertainty principle,
	\textit{Physical Review} \textbf{34} 573-574.
	Reprinted in \cite[127-128]{whi:zur} 

	\bibitem{schr:zum}
	Schr\"odinger, E. (1930),
	The uncertainty relations in quantum mechanics,
	\textit{Zum Heisenbergschen Unsch\"arfeprinzip, 
	Sitzungsberichte der Preussischen Akademie der Wissenschaften, 
	Physikalisch-mathematische Klasse} \textbf{14} 296-303.

	\bibitem{grif:quan}
	Griffiths D. (2005),
	\textit{Quantum Mechanics},
	New Jersey, Pearson.
	
	\bibitem{whi:zur}
	Wheeler, J.A. and W.H. Zurek (eds) (1983),
	\textit{Quantum Theory and Measurement},
	Princeton N.J., Princeton University Press.
\end{thebibliography}
\end{document}